\documentclass[12pt]{emulateapj} 
\newcommand{\Mearth}{$M_\oplus$}

\usepackage{amsmath}
\usepackage{amssymb}
\usepackage{amstext}
\usepackage{color}
\usepackage{graphicx}
\usepackage{epstopdf}
\usepackage{epsfig}
\usepackage{color}
\definecolor{myColor}{rgb}{0.9,0.9,0.9}  
\begin{document}
\renewcommand\bottomfraction{.9}
\shorttitle{Carbon-rich giant planets} 
\title{Carbon-rich Giant Planets: Atmospheric Chemistry, Thermal Inversions, Spectra, and Formation conditions}
\author{Nikku Madhusudhan\altaffilmark{1}, Olivier Mousis\altaffilmark{2}, Torrence V. Johnson\altaffilmark{3} \& Jonathan I. Lunine\altaffilmark{4}} 
\altaffiltext{1}{Department of Astrophysical Sciences, Princeton University, Princeton, NJ 08544, USA {\tt nmadhu@astro.princeton.edu}} 
\altaffiltext{2}{Institut UTINAM, CNRS-UMR 6213, Observatoire de Besan\c con, BP 1615, 25010 Besan\c{c}on Cedex, France}
\altaffiltext{3}{Jet Propulsion Laboratory, California Institute of Technology, Pasadena, CA 91109, USA}
\altaffiltext{4}{Department of Astronomy, Cornell University, Ithaca, NY 14853, USA}

\begin{abstract}
The recent inference of a carbon-rich atmosphere, with C/O $\geq$ 1, in the hot Jupiter WASP-12b motivates the exotic new class of carbon-rich planets (CRPs). We report a detailed study of the atmospheric chemistry and spectroscopic signatures of carbon-rich giant planets (CRGs), the possibility of thermal inversions in their atmospheres, the compositions of icy planetesimals required for their formation via core accretion, and the apportionment of ices, rock, and volatiles in their envelopes. Our results show that CRG  atmospheres probe a unique region in composition space, especially at high temperature ($T$). For atmospheres with C/O $\geq$ 1, and $T \gtrsim$ 1400 K in the observable atmosphere, most of the oxygen is bound up in CO, while  H$_2$O is depleted and CH$_4$ is enhanced  by up to two or three orders of magnitude each, compared to equilibrium compositions with solar abundances (C/O = 0.54). These differences in the spectroscopically dominant species for the different C/O ratios cause equally distinct observable signatures in the spectra. As such, highly irradiated transiting giant exoplanets form ideal candidates to estimate atmospheric C/O ratios and to search for CRPs. We also find that the C/O ratio strongly affects the abundances of TiO and VO, which have been suggested to cause thermal inversions in highly irradiated hot Jupiter atmospheres. A C/O = 1 yields TiO and VO abundances of $\sim$100 times lower than those obtained with equilibrium chemistry assuming solar abundances, at $P \sim 1$ bar. Such a depletion is adequate to rule out thermal inversions due to TiO/VO even in the most highly irradiated hot Jupiters, such as WASP-12b. We estimate the compositions of the protoplanetary disk, the planetesimals, and the envelope of WASP-12b, and the mass of ices dissolved in the envelope, based on the observed atmospheric abundances. Adopting stellar abundances (C/O = 0.44) for the primordial disk composition and low-temperature formation conditions ($T \lesssim 30$ K) for WASP-12b leads to a C/O ratio of 0.27 in accreted planetesimals, and, consequently, in the planet's envelope. In contrast, a C/O ratio of 1 in the envelope of WASP-12b requires a substantial depletion of oxygen in the disk, i.e. by a factor of $\sim$ 0.41 for the same formation conditions. This scenario also satisfies the constraints on the C/H and O/H ratios reported for WASP-12b. If, alternatively, hotter conditions prevailed in a stellar composition disk such that only H$_2$O is condensed, the remaining gas can potentially have a C/O $\sim$ 1. However, a high C/O in WASP-12b caused predominantly by gas accretion would preclude super-stellar C/H ratios which also fit the data.
\end{abstract}

\keywords{planetary systems --- planets and satellites: general --- planets and satellites: individual (WASP-12b)}

\section{Introduction} 

Carbon-rich planets (CRPs) are the exotic new possibility in the repertoire of extrasolar planets. We define a CRP as a planet with a carbon to oxygen (C/O) ratio $\geq 1$. The first carbon-rich atmosphere was inferred recently for the very hot Jupiter WASP-12b (Madhusudhan et al.~2011). Observations of broadband infrared photometry with the {\it Spitzer Space Telescope} (Campo et al. 2010) and from the ground (Croll et al. 2010) were used by Madhusudhan et al.~(2011) to obtain a stringent constraint of C/O $\ge$ 1 in the dayside atmosphere of WASP-12b. The observations revealed a substantial depletion of H$_2$O, and an overabundance of CH$_4$, as compared to a solar abundance chemical equilibrium model. The observations also indicated the absence of a strong thermal inversion (i.e. a `stratosphere') in the dayside atmosphere, which was surprising for the extremely irradiated atmosphere of WASP-12b based on the TiO/VO hypothesis (Hubeny et al. 2003; Fortney et al. 2008). The 
 interior composition of WASP-12b is unknown. However, CRPs are known to posses chemically distinct interiors, atmospheres, and formation conditions (Lodders 2004; Kuchner \& Seager 2005; Bond et al. 2010) from the commonly assumed oxygen-rich planetary compositions, which are based on the solar C/O ratio of 0.5 (Asplund et al. 2005).  Atmospheres of over a dozen transiting exoplanets have now been observed from space and/or from ground (Seager \& Deming 2010). WASP-12b being the first giant planet to have a robust constraint on its atmospheric C/O ratio, one can only wonder about the compositional diversity of extrasolar planets. 

Theoretical studies in the recent past had anticipated the existence of planets with high C/O ratios. The oxygen abundance, 
and hence the value of C/O, of Jupiter is presently unknown (e.g. Atreya et al. 2005). However, based on the lower-limit on O/H measured by the {\it Galileo} probe, Lodders (2004) studied the hypothetical possibility of Jupiter being a CRP. If Jupiter were carbon-rich, she suggested, it could have formed with dominant tarry planetesimals instead of planetesimals dominant in water-ice as expected in the solar system based on the composition of minor bodies in the solar system. Following Lodders (2004), Kuchner and Seager (2005) studied  the possibility of low-mass extrasolar carbon planets ($\lesssim$ 60 M$_\oplus$), characterized by  atmospheres rich in hydrocarbons and deficient in oxygen-rich gases, compared to atmospheres in chemical equilibrium with solar abundances. More recently, Bond et al. (2010) studied terrestrial-mass exoplanets over a wide range of C/O ratios. While metallic silicates dominate the interiors of oxygen-rich terrestrial planets like Earth, those with C/O $\ge$ 0.8 would host interiors that are dominated by carbon-rich solids such as SiC, graphite, and diamond. 

The first detection of a carbon-rich planetary atmosphere, of WASP-12b, opens a new class of exoplanets, with atmospheres, interiors, and/or formation mechanisms, likely very different from expectations based on chemistry with solar abundances. Our focus in this study is on close-in extrasolar giant CRPs whose atmospheres are hydrogen-dominated, i.e. hot Jupiters, hot 
Neptunes, and, possibly, super-Earths\footnote{It is presently unknown if super-Earth atmospheres are hydrogen-rich like that of Neptune or are hydrogen-poor like those of terrestrial planets; the diversity of planetary properties seen to date suggests that both will be found.}. We refer to them as carbon-rich giants (CRGs) in the present work. 

We report a detailed study of atmospheric chemistry, the possibility of thermal inversions, and thermal spectra of close-in CRGs, along with the chemical conditions required for their formation via core accretion. We first present the unique phase-space of molecular abundances in CRG atmospheres and their observable signatures in emergent spectra for planets in different irradiation regimes. Kuchner \& Seager (2005) present a similar analysis for Neptune mass carbon planets. We then address the question of whether thermal inversions can form in carbon-rich hot Jupiters, in order to explain the lack of a strong thermal inversion in WASP-12b. Finally, we use a model based on the core accretion scenario to constrain the primordial conditions required for the formation of WASP-12b, and to estimate the elemental abundances in the planetary envelope, along with the  apportionment of ices, rock, and volatiles in the envelope.

In what follows, we first describe our modeling methods in section~\ref{sec:methods}. We present our results on CRG atmospheres in section~\ref{sec:atmos}. In section~\ref{sec:abund}, we constrain the elemental abundances in the planetary envelope and in the protoplanetary disk required to generate a CRG atmosphere, using WASP-12b as an example. We summarize our conclusions and discuss our results in section~\ref{sec:discussion}.

\section{Methods}
\label{sec:methods}

We employ numerical methods to model three different aspects of carbon-rich giant planets: (a) atmospheric 
chemistry (b) thermal spectra, and (c) composition of planetesimals accreted by the forming giant 
planet. 

\subsection{Atmospheric Chemistry}
\label{sec:methods-chem}

We determine volume mixing ratios of the dominant molecular species using the approach described in 
Madhusudhan \& Seager (2011). We consider hydrogen-dominated atmospheres with temperatures 
typical of known transiting extrasolar giant planets (T$_{\rm eff}$ $\sim$ 700 -- 3000) and with different elemental 
abundances. We compute the molecular mixing ratios generally under the assumption of thermochemical 
equilibrium, considering the effects of non-equilibrium thermochemistry where desirable. We use the equilibrium 
chemistry code originally developed in Seager et al. (2000), and subsequently updated and/or used in several recent 
works (Seager et al. 2005; Kuchner \& Seager 2005; Miller-Ricci et al. 2009; Madhusudhan \& Seager 2011). The code computes gas phase molecular mixing ratios for 172 molecules, resulting from abundances of 23 atomic species, by minimizing the net Gibbs free energy of the system. The multi-dimensional Newton-Raphson method described in White et al. (1958) is used for the minimization. We adopt polynomial fits for the free energies of the molecules based on Sharp \& Huebner (1990). 

For the purpose of the present study, the major chemically and spectroscopically dominant species included in our model 
atmospheres are H$_2$O, CO, CH$_4$, and CO$_2$, in addition to H$_2$ which contributes H$_2$-H$_2$ collision 
induced opacity. Equilibrium mixing ratios of these molecules are computed for different elemental abundances. 
Specifically, we compare the abundances of the major species H$_2$O, CO, CH$_4$ for different C/O ratios, spanning 
oxygen-rich (C/O $<$ 1) and carbon-rich (C/O $>$ 1) compositions over a wide range in pressure-temperature ($P$-$T$) 
space. We then identify the temperature regimes in which C-rich atmospheres are chemically  distinct, and identifiable in 
spectra, from O-rich atmospheres. A similar analysis for two C/O ratios (0.5 and 1.01) and for the specific case of Neptune-mass carbon planets was reported earlier by Kuchner \& Seager (2005). 

The concentrations of species, particularly CO and CH$_4$, can be driven out of equilibrium by transport rapid enough to move material out of a given pressure and temperature regime in a time short compared to the chemical equilibration time  (Yung \& Demore 1999). Vertical mixing is known to be prevalent in atmospheres of brown dwarfs (Noll et al. 1997; Saumon et al. 2003, 2006), giant planets in the solar system (Prinn \& Bashay, 1976; Yung et al. 1988; Fegley and Lodders, 1994; Visscher et al. 2010), and in extrasolar giant planets (Stevenson et al. 2010; Madhusudhan \& Seager 2011). The effects of non-equilibrium chemistry are by their nature more pronounced with decreasing temperature, implying that atmospheres of the cooler population of exoplanets (e.g. GJ~436b, GJ~1214b, HD~189733b, etc.) are more susceptible to CO -- CH$_4$ disequilibrium  (Zahnle et al. 2009, Line et al. 2011, Moses et al. 2011, Miller-Ricci Kempton et al. 2011). Consequently, in this study we investigate the effects of CO -- CH$_4$ disequilibrium due to vertical mixing on the observable CO abundances in some special cases at low temperature. 

The set of reactions governing chemical equilibrium between the dominant carbon-bearing species CH$_4$ and CO is summarized by: 

\begin{equation}
   {\rm CO + 3H_2  \rightleftharpoons CH_4 + H_2O}.
   \label{eq:eddy_1}
\end{equation}

At high pressures and temperatures in the deeper layers of the atmospheres (for $P \gtrsim 1$ 
bar), the reaction timescales are short and the species  approach chemical equilibrium. Higher in the atmosphere where temperatures are lower, turbulent processes can cause timescales of vertical mixing 
to be shorter than the chemical timescales, thereby driving the atmosphere out of equilibrium. 
Our approach for computing the non-equilibrium concentrations of CO and CH$_4$ is described 
in Madhusudhan \& Seager (2011). 

\subsection{Influence of Chemistry on Thermal Inversions}
\label{sec:thermal_inv}

A thermal inversion, like the Earth's stratosphere, is a region in the atmosphere where temperature increases outward (see Madhusudhan \& Seager 2010 for a review). Observations of several hot Jupiters have been interpreted as containing  evidence for thermal inversions in their dayside atmospheres (e.g. Burrows et al. 2008; Knutson et al. 2008, 2009; Machalek et al. 2009; Christiansen et al. 2010; Madhusudhan \& Seager 2009,2010). Thermal inversions in hot Jupiters have been hypothesized as caused by  strong optical absorption due to TiO and VO at high altitudes (Hubeny et al.~2003; Fortney et al.~2008; but cf Zahnle et al. 2008, Knutson et al. 2010). 

Based on equilibrium abundances of TiO and VO, assuming solar elemental abundances, Fortney et al. (2008) suggested that hot Jupiters with high irradiation levels would likely be hot enough to host TiO and VO in gaseous form leading to thermal inversions. On the contrary, Spiegel et al. (2009) showed that TiO/VO, being heavy molecules, are prone to gravitational settling, and would therefore be  deficient at high altitudes in the atmosphere. The amount of TiO required to explain thermal inversions would have to result from vigorous vertical mixing in the atmosphere. Even then, the presence of possible cold traps can cause further depletion of TiO via condensation. VO is much less abundant than TiO. Consequently, TiO/VO might be able to cause inversions in only the extreme end members even amongst the very hot Jupiters of the pM class (Spiegel et al. 2009). However, the recent non-detection of a strong thermal inversion in one of the most highly irradiated exoplanets known, WASP-12b, (Madhusudhan et al. 2011) poses a new challenge to the TiO hypothesis. 

In this work, we investigate the cause behind the absence of a strong thermal inversion in WASP-12b. We compute abundances of TiO and VO in chemical equilibrium over a grid in temperature and pressure space representative of highly irradiated atmospheres. We investigate the influence of C/O ratio on the TiO and VO abundances by computing the TiO and VO abundances at different C/O ratios spanning oxygen-rich and carbon-rich regimes, i.e. solar C/O of 0.54 (as assumed by Fortney et al. 2008) and C/O $\geq 1$. By comparing the TiO and VO abundances obtained in the different regimes, we assess the likelihood of  thermal inversions in high-temperature carbon-rich atmospheres, such as that of WASP-12b, versus those in oxygen-rich atmospheres.  

\subsection{Thermal Spectra}
\label{sec:methods-rt}

We investigate observable signatures of CRG atmospheres over a representative range in effective temperature 
encompassing presently known transiting giant exoplanets. We report model spectra of dayside thermal emission, which can be observed for transiting planets at secondary eclipse. We  compute the model spectra using a 1-D line-by-line radiative transfer code for exoplanetary atmospheres developed in Madhusudhan \& Seager (2009). Given a pressure-temperature ($P$-$T$) profile and the molecular abundance profiles, the code computes the emergent flux by solving the radiative transfer equation along with constraints of LTE, hydrostatic equilibrium, and global energy balance. The model includes the major molecular and continuum opacity sources. We assume H$_2$-rich atmospheres, and we include opacities due to H$_2$O, CO, CH$_4$, CO$_2$, NH$_3$, TiO, VO, and H$_2$-H$_2$ CIA absorption. Our molecular line data are from Freedman et al. (2008), Freedman (personal communication, 2009),  Rothman et al. (2005), Karkoschka \& Tomasko (2010), and Karkoschka (personal communication, 2011). We obtain the H$_2$-H$_2$ collision-induced opacities from Borysow et al. (1997) and Borysow (2002). In order to compute the planet-star flux ratios, we divide the planetary spectrum by a Kurucz model of the stellar spectrum derived from Catelli
  \& Kutucz (2004). 

We compute thermal spectra over a nominal range of irradiation levels (or temperatures), planet types, and C/O ratios. We consider four representative irradiation and planetary regimes: a very hot Jupiter, a hot Jupiter, a hot Neptune, and a warm Neptune. The four chosen planet types have equilibrium temperatures ($T_{\rm eq}$) of 2300 K, 1500 K, 1000 K, and 750 K, and we adopt pressure-temperature ($P$-$T$) profiles motivated by those published in the literature using our present modeling approach (e.g. Madhusudhan \& Seager 2009, 2010, 2011; Madhusudhan et al. 2011). For each case, we adopt the stellar and planetary properties of a known transiting exoplanetary system: hot Jupiter (HD~189733b; Bouchy et al. 2005), very hot Jupiter (HD~189733b with stellar irradiation scaled up), hot Neptune (HAT-P-11b; Bakos et al. 2010), warm Neptune (GJ~436b; Gillon et al. 2007). For each system, we compute mixing ratios of the major molecular species using the prescription described in Section~\ref{sec:methods-chem}, for C/O ratios of solar (0.54), 1, and 3. For the atomic abundances, we fix the O/H ratio at solar (Anders and Grevesse 2005), and we change the C/H ratio to obtain the corresponding C/O ratios. We then compute thermal spectra for each combination of $P$-$T$ profile and composition using our radiative transfer model described above. 

\subsection{Formation Conditions and Composition of Icy Planetesimals}
\label{sec:compo}

Close-in giant planets are thought to have originated in the cold outer regions of protoplanetary disks and migrated inwards until they stopped at closer orbital radii (Goldreich \& Tremaine 1980; Fogg \& Nelson 2005, 2007). Here we assume that the giant planet was formed via core-accretion. We calculate the composition of the icy planetesimals accreted by the forming planet following the approach developed in Mousis et al. (2009b, 2011). In this scenario, building blocks accreted by the protoplanet may have formed all along its radial migration pathway in the protoplanetary disk. However, in this work, we assume that only the planetesimals produced beyond the snow line, i.e. those possessing a significant fraction of volatiles, materially affected the observed O and C abundances due to their vaporization when they entered the envelope of the planet (Mousis et al. 2009b, 2011). This hypothesis is supported by the work of Guillot \& Gladman (2000) who showed that planetesimals 
 delivered to a planet with a mass similar to or greater than that of Jupiter are ejected rather than accreted. 
This mechanism should then prevent further noticeable accretion of solids by the planet during its migration below the snow line. 

We demonstrate our model for the specific case of WASP-12b. The composition of planetesimals accreted by WASP-12b can be inferred from the formation sequence and the composition of the different ices formed beyond the snow line of the protoplanetary disk. This model has been used to interpret the observed volatile enrichments in the atmospheres of Jupiter and Saturn in a way consistent with the heavy element content predicted by interior models (Mousis et al. 2009a) and also the apparent carbon deficiency observed in the Hot Jupiter HD189733b (Mousis et al. 2009b, 2011). It is based on a predefined initial gas phase composition in which elemental abundances reflect those of the host star and describes the process by which volatiles are trapped in icy planetesimals formed in the protoplanetary disk. {Oxygen, carbon, nitrogen, sulfur and phosphorus are postulated to exist only in the form of H$_2$O, CO, CO$_2$, CH$_3$OH, CH$_4$, N$_2$, NH$_3$, H$_2$S and PH$_3$. We set CO/CO$_2 $/CH$_3$OH/CH$_4$ = 70/10/2/1 in the gas phase of the disk, values that are consistent with the ISM measurements considering the contributions of both gas and solid phases in the lines of sight (Frerking et al. 1982; Ohishi et al. 1992; Ehrenfreund  \& Schutte 2000; Gibb et al. 2000). In addition, S is assumed to exist in the form of H$_2$S, with H$_2$S/H$_2$ = 0.5 $\times$ (S/H$_2$)$_{\odot}$, and other refractory sulfide components (Pasek et al. 2005). We also consider N$_2$/NH$_3$ = 1/1 in the disk's gas-phase. This value is compatible with thermochemical calculations in the solar nebula that take into account catalytic effects of Fe grains on the kinetics of N$_2$ to NH$_3$ conversion (Fegley 2000). In the following, we adopt these mixing ratios in our model of the protoplanetary disk. Once the abundances of these molecules have been fixed, the remaining O gives the abundance of H$_2$O. 

\begin{figure}[ht]
\centering
\includegraphics[width = 2.7in, angle=90]{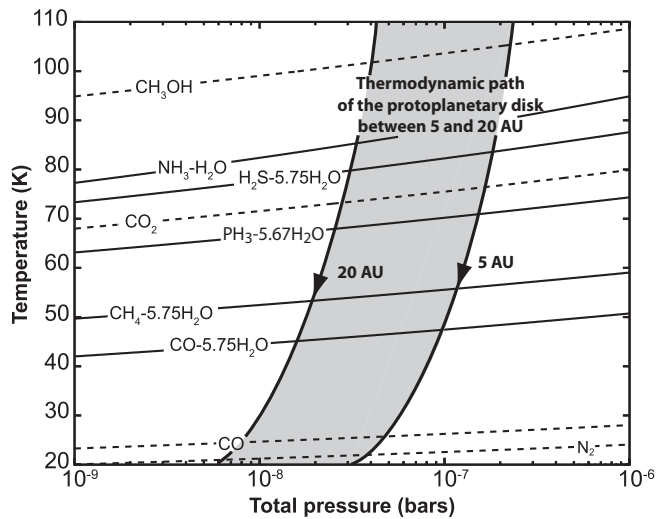}
\caption{Equilibrium curves of ices formed in a protoplanetary disk owning an elemental composition identical to that of WASP-12 and assuming a full efficiency of clathration. Ices include hydrate (NH$_3$-H$_2$O), structure I (X-5.75H$_2$O) and structure II (X-5.67H$_2$O) clathrates (solid lines) and pure condensates (dotted lines). The ensemble of thermodynamic paths (grey area) ranging between 5 and 20 AU in the pressure-temperature space of the disk. Species remain in the gas phase above the equilibrium curves. Below, they are trapped as clathrates or simply condense.}
\label{fig:cool1}
\end{figure}
 
The process of volatile trapping in planetesimals, illustrated in Fig.~\ref{fig:cool1}, is calculated using the equilibrium curves of hydrates, clathrates and pure condensates, and the ensemble of thermodynamic paths detailing the evolution of temperature and pressure in the 5--20 AU range of the protoplanetary disk. We refer the reader to the works of Papaloizou \& Terquem (1999) and Alibert et al. (2005b) for a full description of the turbulent model of accretion disk used here. This model postulates that viscous heating is the predominant heating source, assuming that the outer parts of the disk are protected from solar irradiation by shadowing effect of the inner disk parts. In these conditions, temperature in the planet-forming region can decrease down to very low values ($\sim$20 K; Mousis et al. 2009a).
 
For each ice considered in Fig.~\ref{fig:cool1}, the domain of stability is the region located below its corresponding equilibrium curve. The clathration process stops when no more crystalline water ice is available to trap the volatile species. The equilibrium curves of hydrates and clathrates derive from the compilation of published experimental work by Lunine \& Stevenson (1985), in which data are available at relatively low temperatures and pressures. On the other hand, the equilibrium curves of pure condensates used in our calculations derive from the compilation of laboratory data given in the CRC Handbook of Chemistry and Physics (Lide 2002). For each ice considered, the domain of stability is the region located below its corresponding equilibrium curve. The clathration process stops when no more crystalline water ice is available to trap the volatile species. Note that, in the pressure conditions of the solar nebula, CO$_2$ is the only species that crystallizes at a higher temperature than its associated clathrate. We then assume that solid CO$_2$ is the only existing condensed form of CO$_2$ in this environment. In addition, we have considered only the formation of pure ice of CH$_3$OH in our calculations since, to our best knowledge, no experimental data concerning the equilibrium curve of its associated clathrate have been reported in the literature. The intersection of a thermodynamic path at a given distance from the star with the equilibrium curves of the different ices allows determination of the amount of volatiles that are condensed or trapped in clathrates at this location in the disk. Indeed, the volatile, $i$, to water mass ratio in these planetesimals is determined by the relation (Mousis \& Gautier 2004; Mousis \& Alibert 2006):         

\begin{equation}
{m_i = \frac{X_i}{X_{H_2O}} \frac{\Sigma(r; T_i, P_i)}{\Sigma(r; T_{H_2O}, P_{H_2O})}},
\end{equation}

\noindent where $X_i$ and $X_{H_2O}$ are the mass mixing ratios of the volatile $i$ and H$_2$O with respect to H$_2$ in the nebula, respectively. $\Sigma(R; T_i, P_i)$ and $\Sigma(R; T_{H_2O}, P_{H_2O})$ are the surface density of the nebula at a distance $r$ from the Sun at the epoch of hydratation or clathration of the species $i$, and at the epoch of condensation of water, respectively. From ${\it m_i}$, it is possible to determine the mass fraction $M_i$ of species $i$ with respect to all the other volatile species taking part to the formation of an icy planetesimal via the following relation:

\begin{equation}
{M_i = \frac{m_i}{\displaystyle \sum_{j=1,n} m_j}},
\end{equation}

\noindent with $\displaystyle \sum_{i=1,n} M_i = 1$.\\

Assuming that, once condensed, the ices add to the composition of planetesimals accreted by the growing planet along its migration pathway, this allows us to reproduce the volatile abundances by adjusting the mass of planetesimals that vaporized when entering the envelope. Note that, because the migration path followed by the forming WASP-12b is unknown, the 5--20 AU distance range of the ensemble of thermodynamic paths has been arbitrarily chosen to determine the composition of the accreted ices. The adoption of any other distance range for the planet's path beyond the snow line would not affect the composition of the ices (and thus WASP-12b global volatile abundances) because it remains almost identical irrespective of i) their formation distance and ii) the input parameters of the disk, provided that the initial gas phase composition is homogeneous (Marboeuf et al. 2008).

\begin{figure*}[ht]
\centering
\includegraphics[width = \textwidth]{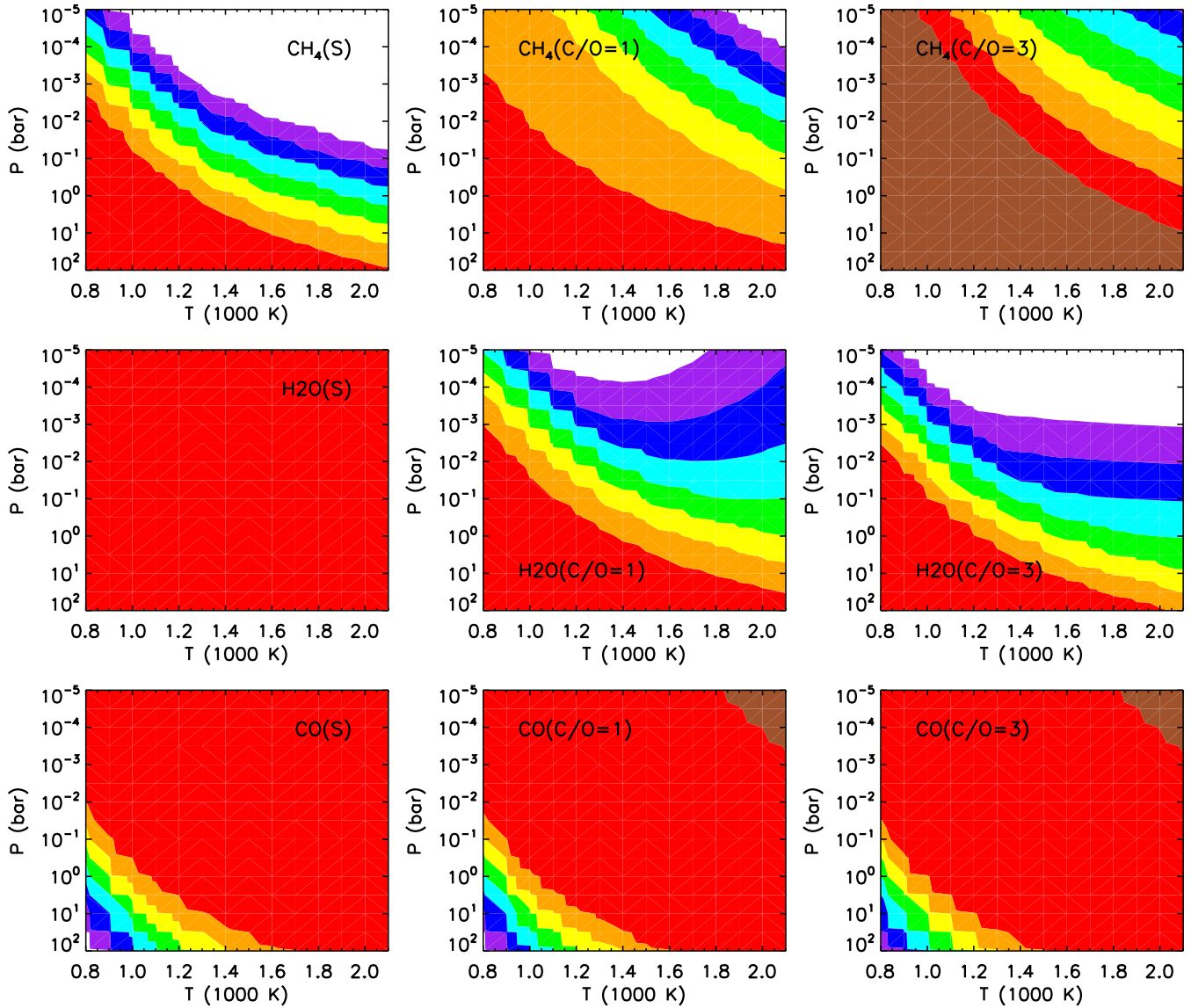}
\caption{Concentrations of CH$_4$, H$_2$O, and CO, predicted by equilibrium chemistry assuming C/O ratios of solar (or 0.54, denoted by `S'), 1, and 3. Each panel shows contours of molecular mixing ratios (i.e. ratio by number density) with respect to molecular hydrogen, in pressure-temperature space. The left, center, and right panels show mixing ratios for C/O of solar, 1.0 and 3.0, respectively. The brown, red, orange, yellow, green, cyan, blue and purple contours correspond to mixing ratios greater than $10^{-3}$, $10^{-4}$, $10^{-5}$, $10^{-6}$, $10^{-7}$, $10^{-8}$, $10^{-9}$, and $10^{-10}$, respectively. Higher C/O ratios predict higher CH$_4$ and lower H$_2$O, compared to those predicted by solar C/O.}
\label{fig:equib}
\end{figure*}

\section{Results: Atmospheres}
\label{sec:atmos}

In this section, we report on the atmospheric phase space and spectroscopic signatures of CRGs. We begin with 
an overview of molecular abundances of the major species expected in equilibrium, along with potential 
deviations that could be caused by non-equilibrium chemistry in the low temperature regime. We then 
investigate whether TiO and VO could cause thermal inversions in CRGs. Finally, we present model spectra of 
carbon-rich atmospheres over different irradiation regimes vis-a-vis those with solar abundance 
atmospheres. 

\subsection{Atmospheric Chemistry}
\label{sec:atmos-chem}

\begin{figure*}[ht]
\centering
\includegraphics[width = \textwidth]{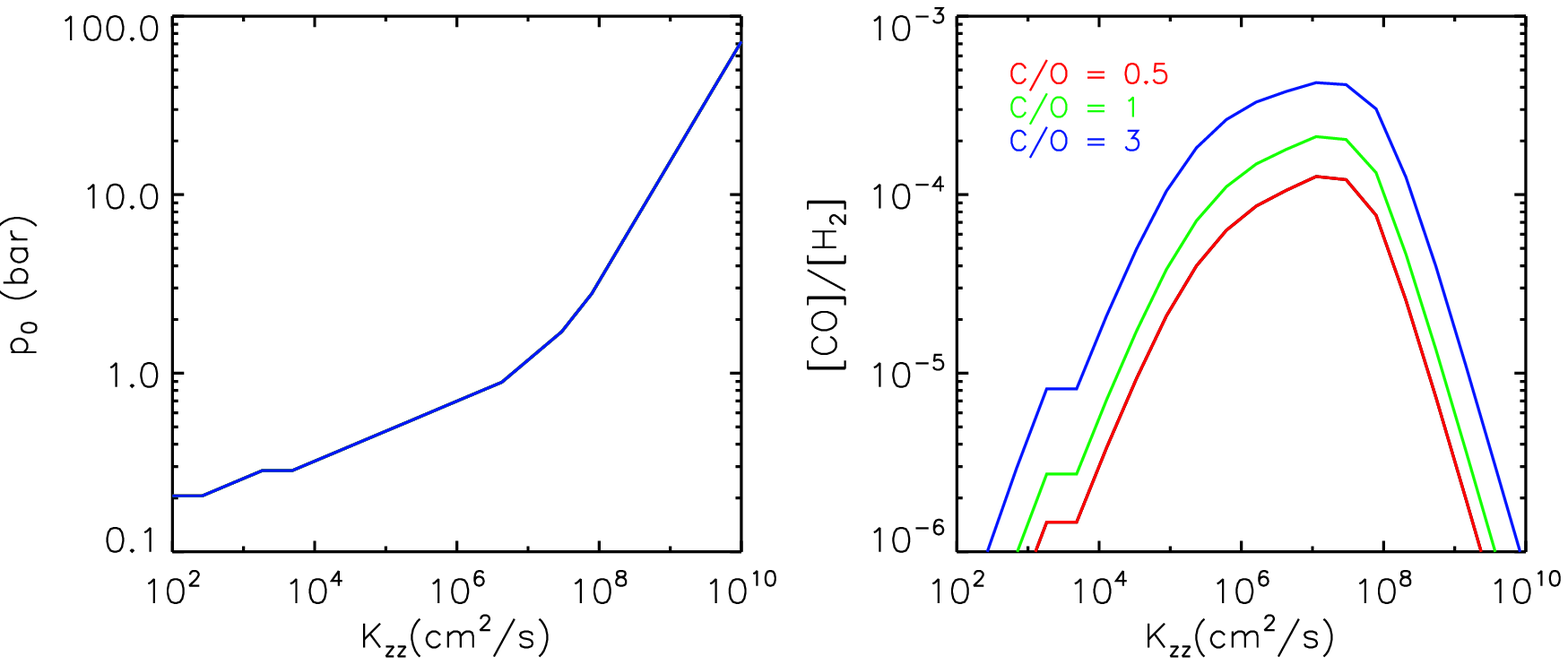}
\caption{Non-equilibrium CO-CH$_4$ thermochemistry due to vertical mixing in GJ~436b for different C/O ratios. We follow the approach of  Madhusudhan \& Seager (2011) and use the atmospheric temperature profile of GJ~436b reported therein. The left panel shows the dependence of the quench pressure (p$_0$) on the eddy diffusion coefficient, K$_{zz}$. Higher  K$_{zz}$ allows vertical mixing from deeper layers of the atmosphere which are at higher pressures; the curves overlap for all C/O ratios.  The right panel shows the dependence of the quenched CO abundance on K$_{zz}$. Significantly high values of CO, between (1 - 5) $\times 10^{-4}$ are attainable for K$_{zz} \sim 10^7$ cm$^2$/s. However, for any K$_{zz}$, the differences between observable CO abundances for the different C/O ratios are only marginal (see discussion at end of section~\ref{sec:atmos-chem}). }
\label{fig:eddy}
\end{figure*}

Atmospheric chemistry in hydrogen-dominated atmospheres is critically influenced by the ratio of carbon to oxygen. 
The major species resulting from carbon-oxygen chemistry in chemical equilibrium are 
water vapor (H$_2$O), carbon monoxide (CO), and methane (CH$_4$). For solar abundances (C/O = 0.54), 
CH$_4$ is the dominant carrier of carbon at low temperatures ($T$), and CO is the dominant carbon carrier 
at high $T$, as shown in Fig.~\ref{fig:equib}. The temperature of transition ($T_{\rm CO - CH_4}$) depends 
on the pressure ($P$); at a representative $P \sim 1$ bar, $T_{\rm CO - CH_4} \sim $ 1200--1400 K.  On the 
other hand, H$_2$O is a major carrier of oxygen at all temperatures. At low $T$, almost the entire O is 
contained in H$_2$O, whereas at high $T$, a significant amount of O is also contained in the abundant CO. 
As CO overtakes CH$_4$ at $T \sim T_{\rm CO - CH_4}$, O is present in nearly equal proportions between 
H$_2$O and CO. This distribution of H$_2$O, CO, and CH$_4$ for solar abundances is substantially 
perturbed when the C/O ratio is increased. 

At high T, hydrogen-dominated carbon-rich (with C/O $\geq$ 1) atmospheres bear manifestly different concentrations 
of the major species from those in solar abundance atmospheres, as shown in Fig.~\ref{fig:equib}, and previously suggested by Kuchner \& Seager (2005). One of the most apparent changes is a general enhancement in CH$_4$ 
abundance and depletion in H$_2$O abundance over a wide region in $P$--$T$ space. Most of the O is occupied by 
CO over a wide temperature range. The degree of H$_2$O depletion over solar abundance concentrations increases with T. Considering the C/O = 1 case in Fig.~\ref{fig:equib}, for instance, at a nominal $P \sim 1$ bar, H$_2$O can be depleted 
by a factor of 10-100 for $T$ between 1400 - 2000 K. At lower pressures, such as $P \sim 0.1$ bar, which 
are typically probed by the observations (Madhusudhan et al. 2011), the depletion is even greater. On the contrary, 
over the same temperature and pressure range, the CH$_4$ abundance exceeds that in solar abundance by factors  as high as $10^3$. 

The depletion in H$_2$O and enhancement in CH$_4$, at high T, increase further as the C/O ratio increases beyond 1, as shown for C/O = 3 in Fig.~\ref{fig:equib}. Thus, whereas solar abundances typically suggest H$_2$O mixing ratios of $\sim 5 \times 10^{-4}$ at all temperatures, high C/O ratios can cause H$_2$O mixing ratios as low as $\sim 10^{-7} - 10^{-6}$ in the observable atmospheres ($P \lesssim 0.1 - 1$ bar) at high T ($\gtrsim 1400$ K). Similarly, while CH$_4$ is generally presumed to be scarce at high T, high C/O ratios can yield observable CH$_4$ abundances up to $\sim 10^{-5} - 10^{-4}$. 

The strong influence of the C/O ratio on atmospheric composition at high T indicates that highly irradiated transiting 
hot Jupiters and hot Neptunes are ideal probes of C/O ratios and form good candidates to search for CRPs. 
At lower temperatures ($T \lesssim 1200$ K at $P \sim 1$ bar), the distinction between the C-rich and O-rich regimes are less apparent, as shown in Fig.~\ref{fig:equib}. In this regime, irrespective of the C/O ratio, most of the carbon is carried by CH$_4$, CO is scarce, and H$_2$O is the primary carrier of oxygen. Thus, for C/O = 1, CH$_4$ and H$_2$O are equally abundant, with CH$_4$ being only a factor of 2 more abundant than in a solar abundance model (C/O = 0.54). Such small deviations of  less than a factor of $\sim$10 in molecular abundances are well below the uncertainties in current molecular abundance  determinations for exoplanetary atmospheres (e.g. Swain et al. 2008, Madhusudhan \& Seager 2009, 2011, Madhusudhan et al. 2011. Consequently, based on equilibrium chemistry alone, reliable determination of C/O ratios of cooler giant planets such as GJ~436b, with $T \lesssim 1200$ K at $P \sim 1$ bar, will need high resolution spectra and high precision retrieval of CH$_4
 $ and H$_2$O abundances along with those of other species. 

The concentrations of CO and CH$_4$ are also influenced by non-equilibrium chemistry in the form of vertical mixing, as discussed in section~\ref{sec:methods-chem}, especially at low T. In particular, enhanced CO abundances over that 
possible in chemical equilibrium can be transported from the lower hotter regions of the atmosphere to upper observable 
regions of the atmosphere. As shown in Madhusudhan \& Seager (2010) for GJ~436b, mixing ratios of order $10^{-3}$ are attainable for high metallicities, corresponding to $\sim 10 - 30 \times$ solar. However, 
in the present context the question is whether vertical mixing is sensitive to the overall atmospheric C/O value in the low T regime. Fig.~\ref{fig:eddy} shows the range of CO abundances possible in the observable 
atmosphere of GJ~436b due to non-equilibrium chemistry with different C/O ratios. We find that the differences between 
the CO abundances resulting from the variation in C/O ratios merely reflect the higher C/H abundances in the various 
cases. Note that we fix the O/H and vary the C/H to obtain different C/O ratios. As such, the CO abundance from our  
C/O = 3 case, for example, can also be obtained by a C/O = 0.5 model with $\sim 6$ times higher bulk metallicity.
We therefore find that even with the effect of vertical mixing, high and low C/O atmospheres would be challenging to 
distinguish observationally. In principle, other mechanisms of non-equilibrium chemistry such as photochemistry 
(e.g. Line et al. 2011, Moses et al. 2011) might give rise to effects that could distinguish between the different C/O ratios. 
For example, if methane were destroyed and CO enhanced by non-equilibrium processes (Madhusudhan \& Seager, 2011), 
the observable concentrations of H$_2$O and CO, along with other photochemical products, might be used to estimate the C/O ratio of the atmosphere. 

\begin{figure*}[ht]
\centering
\includegraphics[width = 0.8\textwidth]{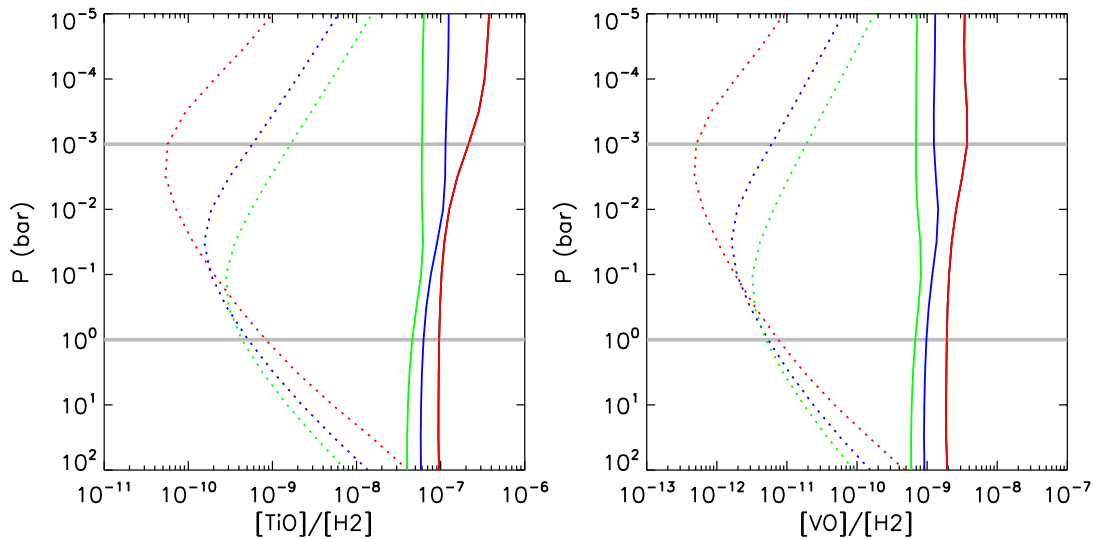}
\caption{ Mixing ratios of TiO and VO for different C/O ratios. The solid and dotted curves show mixing ratios  for C/O ratios of 0.5  and 1.0, respectively. The red, blue, and green colors correspond to isothermal temperature profiles at 2650K, 3000K, and 3250 K. In equilibrium, a C/O $\geq 1$ leads to $\sim 100\times$ lower TiO and VO abundances compared with those obtained from solar abundances. See Section~\ref{sec: atmos-inv}.}
\label{fig:tio}
\end{figure*}

\subsection{Likelihood of Thermal Inversions via the TiO/VO Hypothesis}
\label{sec: atmos-inv}

Our results indicate that carbon-rich atmospheres are less likely to host thermal inversions due to TiO and 
VO. Fig.~\ref{fig:tio} shows molecular mixing ratios of TiO and VO for different C/O ratios over a range of 
temperatures characteristic of highly irradiated hot Jupiters. WASP-12b, which is one of the most extremely 
irradiated hot Jupiters is known to have a lower atmosphere consistent with an isotherm of $\sim$ 3000 K 
(Madhusudhan et al. 2011). We, therefore, consider three isothermal temperature profiles at 2650 K, 3000 K, and 3250 K spanning a representative range for very hot Jupiters. For each temperature, we determine TiO and VO abundances in chemical equilibrium for C/O = 0.54, i.e., solar (solid curves) and C/O = 1 (dotted curves), as shown in 
Fig.~\ref{fig:tio}. We find that a C/O = 1 leads to TiO and VO abundances of $\sim$ 100 times lower than 
the corresponding abundances for solar C/O, at the $P \sim 1$ bar level. The depletion is even more enhanced 
at the 0.1 bar level, where the planetary `photosphere' typically lies.  

The depletion of TiO and VO due to an increased C/O ratio offers a natural explanation for the 
lack of a thermal inversion in the atmosphere of WASP-12b which is known to have a C/O $\geq 1$ 
(Madhusudhan et al. 2011). The $\sim$100 times depletion in TiO and VO for a C/O = 1 as discussed above 
may well be a lower-limit. Considering higher C/O ratios, as are possible in WASP-12b, would lead to 
even greater TiO/VO depletion. Considering non-equilibrium gravitational settling, as suggested to be 
important by Spiegel et al. (2009), could contribute additional depletion. Spiegel et al. (2009) further 
suggest that a depletion in TiO below solar abundances by a factor of $\sim$ 10 is enough to rule out a 
thermal inversion in the atmosphere. Thus, considering all the effects discussed above, we conclude that 
the degree of TiO/VO depletion caused by a C/O $\geq 1$ is adequate to rule out a thermal inversion due 
to TiO and VO even in the most highly irradiated hot Jupiters, such as WASP-12b. It remains to be seen 
if other compounds might exist in highly irradiated carbon-rich atmospheres that could provide visible 
or ultra-violet absorption comparable to TiO/VO, such as ZrO and/or YO found in carbon-rich stars 
(e.g. J\o rgensen 1994) or higher order hydrocarbons. Finally, the abundances 
of TiO/VO and other inversion causing compounds may also be correlated with the stellar activity 
(Knutson et al. 2010).

\begin{figure*}[ht]
\centering
\includegraphics[width = \textwidth]{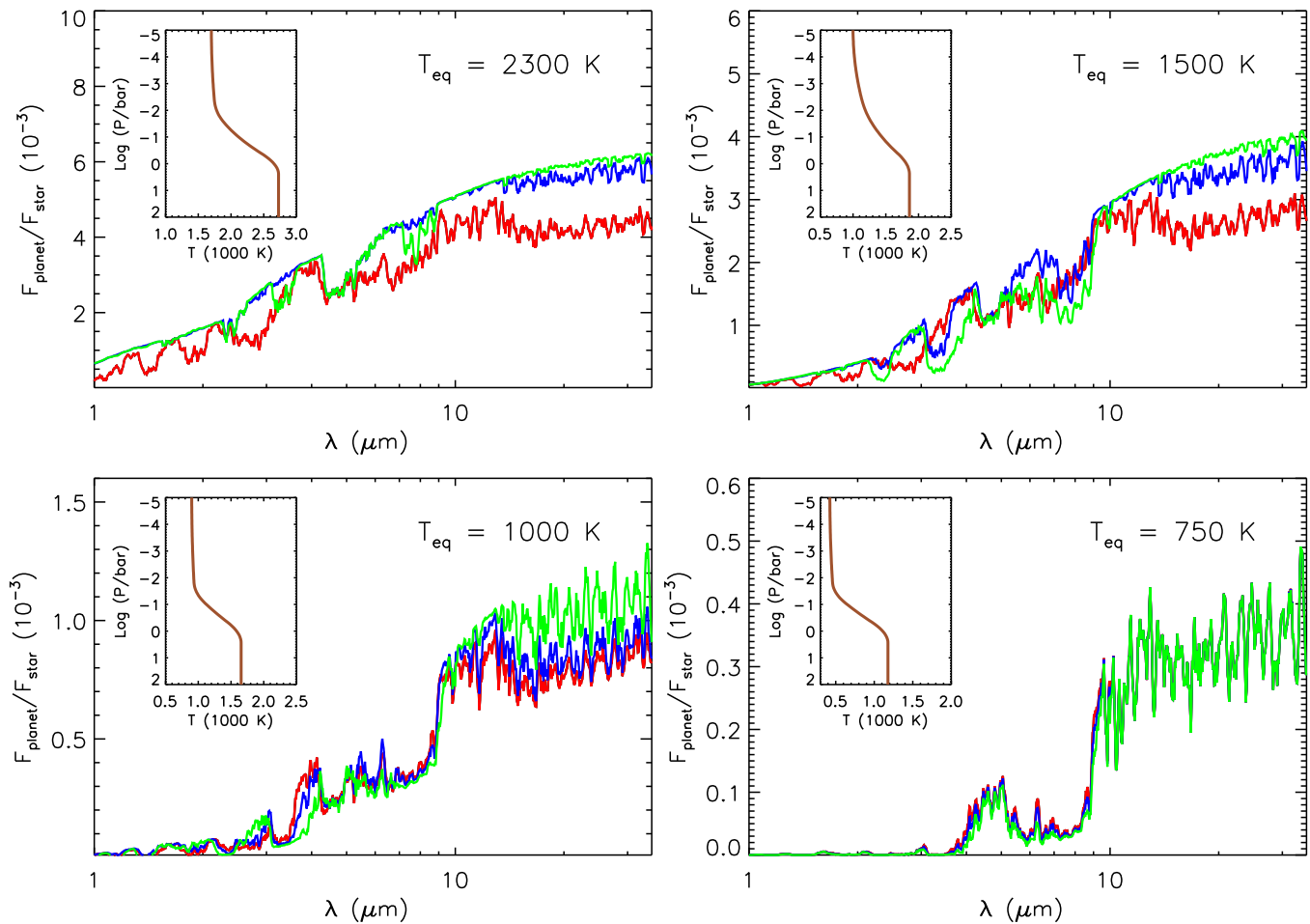}
\caption{Model Spectra of carbon-rich planets. Thermal spectra are shown for giant planets with four different equilibrium temperatures ($T_{\rm eq}$), signifying different irradiation levels The pressure-temperature profiles are shown in the inset. The red curves show models with solar composition, and the blue and green curves show models with C/O of 1 and 3, respectively. The corresponding pressure-temperature profiles are shown in the insets.} 
\label{fig:spectra}
\end{figure*}

\subsection{Spectroscopic Signatures}

The chemical compositions of CRGs lead to observable spectroscopic signatures over a wide 
range of atmospheric temperature profiles. The key differences between spectra of CRGs and 
those of giant planets with solar C/O ratio result from their differing concentrations of the major 
molecular absorbers, H$_2$O, CO and CH$_4$, depending on the temperature regime. As 
discussed in section~\ref{sec:atmos-chem}, at high temperatures ($T \gtrsim$ 1400 K, for $P$ $<$ 1 bar), 
giant planet atmospheres with solar C/O ratio (i.e oxygen-rich) are expected to be dominated by 
CO and H$_2$O as the major carbon and oxygen bearing species, while CH$_4$ is scarce. On the 
other hand, atmospheres of CRPs (C/O $\geq 1$) at the same high temperatures are characterized 
by a substantial enhancement of CH$_4$ and depletion of H$_2$O, compared to those with solar C/O; 
the amount of CO remains comparable. At low temperatures, however, the distinction between 
the different C/O regimes are less pronounced. Atmospheres at lower $T_{eq}$ for both C/O regimes 
have comparable amounts of both CH$_4$ and H$_2$O, and much less CO. These compositional 
characteristics of high and low C/O atmospheres in the different temperature regimes are manifested 
in the corresponding spectra. 

As discussed in section~\ref{sec:methods-rt}, we construct dayside thermal spectra of giant exoplanetary 
atmospheres over a representative range of irradiation levels (or temperatures) and C/O ratios. Model spectra 
for the different cases are shown in Fig.~\ref{fig:spectra}. As shown in the top left panel, for an extremely  irradiated hot Jupiter (with $T_{\rm eq} \sim 2300$ K), the spectrum of a carbon-rich atmosphere is distinctly different from one with a solar C/O, across a broad spectral range in the near to mid infrared. In particular, the CRG spectrum  contains high absorption due to the strong CH$_4$ bands around 3.5 $\micron$ and 8 $\micron$, along with milder features in the K band (around 2.2 $\micron$), while predicting high flux (or low absorption) at wavelengths associated with strong H$_2$O absorption (e.g. at wavelengths 1.8 $\micron$, 2.8 $\micron$, 5 - 7 $\micron$, and $>$15 $\micron$). On the contrary, spectra with solar C/O ratios are saturated with absorption features due to H$_2$O, and display minimal CH$_4$ absorption. In both high and low C/O cases, however, the spectra show comparable absorption in the CO bands, around 4.5 $\micron$. The top-right panel of Fig.~
 \ref{fig:spectra} shows thermal spectra of a typical hot-Jupiter with $T_{\rm eq} \sim 1500$ K and with different C/O ratios. The differences between the solar and high C/O cases are still evident across the CH$_4$ and H$_2$O bands. The bottom panels of Fig.~\ref{fig:spectra} show two cases of giant planets at the cooler end of known transiting planets: a hot Neptune with $T_{\rm eq} \sim 1000$ K (bottom-left panel) and a warm Neptune with $T_{\rm eq} \sim 750$ K (bottom-right panel). In both cases, the differences between spectra of models with solar C/O and those with C/O = 1 are much less discernible compared to the hotter cases discussed above. However, spectra of CRGs with $T_{\rm eq} \sim 1000$ K are still distinguishable from solar composition models for CRG atmospheres with C/O $\gtrsim 3$. On the other hand, CRGs with $T_{\rm eq} \sim 750$ K are nearly indistinguishable from solar models even for high C/O ratios.  

Currently known transiting giant exoplanets provide a rich sample for future observational campaigns in search of CRGs. The higher temperature hot Jupiters and hot Neptunes (with $T_{eq} \gtrsim 1000$ K) provide the most favorable conditions to detect CRGs. Such atmospheres show clear spectroscopic differences between solar and carbon-rich compositions, with a signal clearly detectable with existing or forthcoming instruments and analysis techniques (e.g. Madhusudhan et al. 2011). For instance, in a CRG atmosphere observed with the equivalent of four channels of Spitzer IRAC photometry, the thermal spectrum would display a high flux in the 5.8 $\micron$ channel (due to low water absorption) and low 3.6 $\micron$ and 8 $\micron$ fluxes (due to strong methane absorption). On the other hand, giant planets with low irradiation levels, $T_{eq} \lesssim 800$ K, are not  readily distinguishable between the low and high C/O cases, unless high resolution spectroscopy is available allowing high precision retrieval of H$_2$O and CH$_4$ abundances. 

\section{Results: Elemental abundances in WASP-12b}
\label{sec:abund}

In this section, we constrain the elemental abundances required in the protoplanetary disk to explain a C/O $\geq$ 1 in the post--formation planetary envelope , and, hence, in the observable atmosphere. We consider two different compositions for the initial gas phase abundances in our protoplanetary disk model, as shown in Table~\ref{gas}. In both cases, the elemental C, N and S abundances, relative to hydrogen, were adopted from the composition of the host star,  WASP-12, obtained by Fossati et al. (2010), which has C/O = 0.44 (also see Petigura \& Marcy 2011). Because the stellar P abundance is unknown, we have assumed it to be of solar abundance (Asplund et al. 2009) weighted by a factor equal to the average enhancement of the C, N and S abundances in the star relative to those in the sun. Our choice of P enhancement is justified at the precision of current data, since the enhancements of the stellar C, N, and S abundances relative to solar values reported by Fossati et al. (2010) are roughly consistent with each other to within the 1-sigma error bars.

The two cases we consider are different in the oxygen abundance adopted for the primordial disk. In the first case, as shown in Table~\ref{gas}, we have assumed that the O/H in the disk corresponds to the stellar value (Fossati et al. 2010). The formation sequence of the different ices follows the scheme depicted in Fig. \ref{fig:cool1} and shows that the icy component of the planetesimals is essentially composed of a mix of pure condensates and clathrates. In this case, we find that the resulting C/O ratio in the planetesimals, and hence in the planetary envelope after planet formation, is 0.27, which is not consistent with the measured C/O ratio of $\ge 1$ in the atmosphere. In the second case, we have adopted a substellar O abundance (lower by 41\%) in the gas phase of the disk and we find that this assumption allows us to retrieve a composition of planetesimals that matches the observed C/O ratio in WASP-12b. In this case, because the oxygen abundance is strongly depleted compared to case 1, oxygen in the icy part of the planetesimals is entirely distributed among the carbon bearing species and water-ice is absent, as shown in Fig.~\ref{fig:cool2}.

\begin{table}
\caption[]{Elemental and molecular abundances relative to H$_2$ in the protoplanetary disk.}
\begin{center}
\begin{tabular}{lccc}
\hline
\hline
\noalign{\smallskip}
Species X 	&  Case 1				&							& Case 2				\\	
\noalign{\smallskip}
\hline
\noalign{\smallskip}
C			& 					& $7.1 \times 10^{-4}$			& 					\\
N   			& 					& $2.2 \times 10^{-4}$			    					\\
O			& $1.6 \times 10^{-3}$	&							&  $6.6 \times 10^{-4}$	\\
S        		& 					&$3.3 \times 10^{-5}$			&					\\
P			& 					&$7.3 \times 10^{-7}$			&					\\
H$_2$O		& $8.1 \times 10^{-4}$	&							& 0					\\
CO			& $6.0 \times 10^{-4}$	&							& $5.0 \times 10^{-4}$	\\
CO$_2$		& $8.6 \times 10^{-5}$	&							& $7.1 \times 10^{-5}$	\\
CH$_3$OH	& $1.7 \times 10^{-5}$	&							& $1.4 \times 10^{-5}$	\\
CH$_4$		& $8.6 \times 10^{-6}$	&							& $1.3 \times 10^{-4}$	\\

N$_2$		& 					& $7.3 \times 10^{-5}$			&					\\
NH$_3$		& 					& $7.3 \times 10^{-5}$			&					\\
H$_2$S		& 					& $1.6 \times 10^{-5}$			&					\\
PH$_3$		& 					& $7.3 \times 10^{-7}$			&					\\

\hline
\end{tabular}
\end{center}
\tablecomments{Elemental abundances are adopted from Fossati et al. (2010). For Case 2, oxygen is depleted by 41\% (see Section~\ref{sec:abund}).}
\label{gas}
\end{table}

\begin{table}
\caption[]{Composition of icy planetesimals (wt\%).}
\begin{center}
\begin{tabular}{lcc}
\hline
\hline
\noalign{\smallskip}
Species X 	&  Case 1				& Case 2						\\	
\noalign{\smallskip}
\hline
\noalign{\smallskip}
H$_2$O		& $5.6 \times 10^{-1}$	&	0						\\
CO			& $2.5 \times 10^{-1}$	&	$4.7 \times 10^{-1}$			\\
CO$_2$		& $9.7 \times 10^{-2}$	&	$2.2 \times 10^{-1}$			\\
CH$_3$OH	& $1.7 \times 10^{-2}$	&	$3.8 \times 10^{-2}$			\\
CH$_4$		& $2.9 \times 10^{-3}$	&	$7.9 \times 10^{-2}$			\\
N$_2$		& $2.1 \times 10^{-2}$	&	$5.8 \times 10^{-2}$			\\
NH$_3$		& $3.5 \times 10^{-2}$	&	$9.6 \times 10^{-2}$			\\
H$_2$S		& $1.5 \times 10^{-2}$	&	$3.7 \times 10^{-2}$			\\
PH$_3$		& $6.1 \times 10^{-4}$	&	$1.3 \times 10^{-3}$			\\
\hline
\end{tabular}
\end{center}
\label{wt}
\end{table}
 
\begin{figure}
\centering
\includegraphics[width = 2.7in, angle=90]{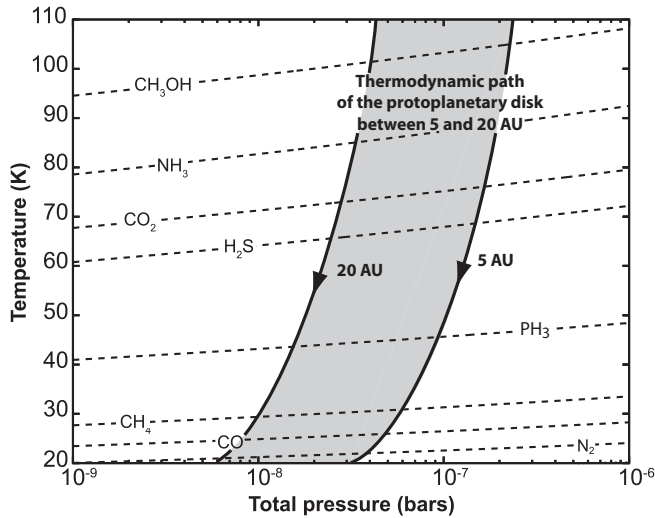}
\caption{Same as in Fig.\ref{fig:cool1} but for a disk owning the same elemental composition as in WASP-12, except the O abundance which is now 41\% of the WASP-12 value. In this case, water does not exist in the disk and only pure condensates form.}
\label{fig:cool2}
\end{figure}
 
\begin{figure}
\centering
\includegraphics[width = 2.7in, angle=0]{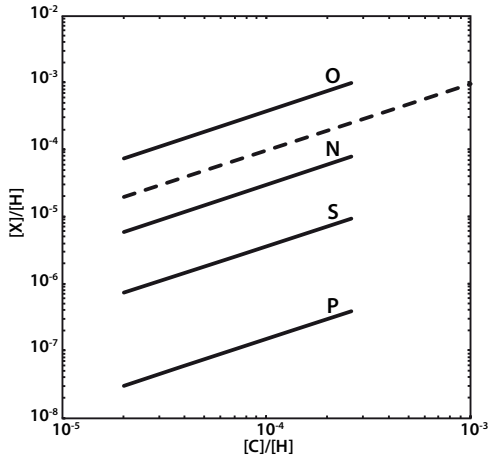}
\caption{Volatile X abundance as a function of the carbon abundance calculated in the atmosphere of WASP-12b. The dashed line corresponds to the range of C and O abundances measured in the planet's atmosphere.}
\label{fig:pred1}
\end{figure}

\begin{figure}
\centering
\includegraphics[width = 2.7in, angle=0]{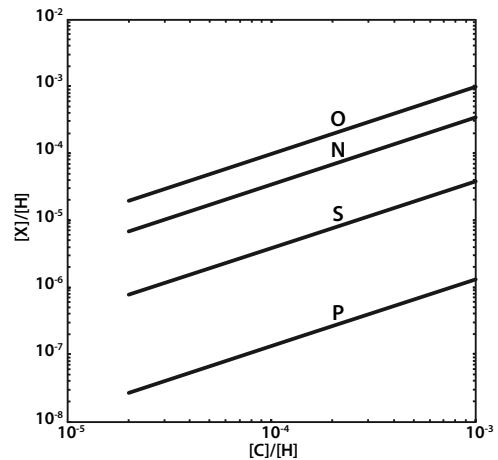}
\caption{Same as Fig. \ref{fig:pred1} but for an O abundance 0.41 times lower than in WASP-12 in the protoplanetary disk. The dashed line corresponding to the range of C and O abundances measured in the planet's atmosphere is now superimposed with the calculated values.}
\label{fig:pred2}
\end{figure}

Table \ref{wt} gives the average composition of planetesimals formed in the protoplanetary disk in the two cases, following the approach depicted in Sec.~\ref{sec:compo}. In case 1, the composition of ices in the planetesimals is dominated by H$_2$O and CO, and is close to that expected from planetesimals formed from a solar-composition gas phase. In case 2, the composition of ices in the planetesimals is dominated by carbon-bearing compounds, with CO, CO$_2$ and CH$_4$ being the three dominant species. This fraction of the planetesimals accreted by the forming hot Jupiter and dissolved into the envelope provides its content in heavy elements. By varying the mass of planetesimals accreted and dissolved within the planet's envelope, our compositional calculations allow retrieval of the corresponding abundances ranges of heavy elements.

Figures~\ref{fig:pred1} and \ref{fig:pred2} present the O, N, S and P abundances calculated as a function of the C abundance in the atmosphere of WASP-12b in cases 1 and 2, respectively. These ranges of abundances have been fitted to the observed C and O abundance ranges (C/H = 2.0 $\times$ 10$^{-5}$--1.0 $\times$ 10$^{-3}$ and O/H = 2.0 $\times$ 10$^{-5}$--1.0 $\times$ 10$^{-3}$) retrieved in the atmosphere of WASP-12b by Madhusudhan et al. (2011). Because C and O measurements in WASP-12b are strongly correlated (Madhusudhan et al. 2011), it is important that the fitted calculations follow the C/O ratio measured in the planet itself. Figure \ref{fig:pred1} shows that case 1 does not reproduce this trend because the C abundance calculated as a function of O abundance is not consistent with the observed correlation between the C and O abundances. On the other hand, Figure \ref{fig:pred2} shows that the adopted substellar O abundance in case 2 permits C/O = 1 in the planet's atmosphere.

The variation of elemental abundances calculated in WASP-12b translates into a mass of ices dissolved in the envelope ranging between 0.8 and 9.7 \Mearth~(including 0.4 to 5.4 \Mearth~of water) in case 1, and between 0.3 and 15.9 \Mearth~(with 0 \Mearth~of water) in case 2. 

\section{Predictions} 

The elemental abundances in the envelope predicted in Fig.~\ref{fig:pred2} lead to an atmospheric 
composition distinctly different from the assumption of solar abundances. Table~\ref{tab:predictions} 
shows mixing ratios of some of the dominant molecular species of C, O, N, S, P, Ti, and V, in 
chemical equilibrium, assuming different elemental abundances, and in the temperature regime of WASP-12b. 
The two molecular compositions in Table~\ref{tab:predictions} assume elemental abundances for 
the two cases, solar and carbon-rich, predicted in Fig.~\ref{fig:pred2}; for the CRP case, we nominally use the solution 
with the maximum O/H. As discussed in section~\ref{sec:atmos-chem}, one of the major differences between the two cases 
is in the concentrations of H$_2$O and CH$_4$. The predictions for the CRP case yield over two orders of 
magnitude lower H$_2$O and over three orders of magnitude higher CH$_4$ compared to those predicted 
assuming solar abundances, as has been observed in Madhusudhan et al. (2011). Beyond the C and O 
species that have already been inferred in WASP-12b, we predict the concentrations of several other species in 
WASP-12b, observations of which could test the elemental abundances predicted by our formation model. 

One of our key predictions is that CRP atmospheres might be characterized by a high content of hydrocarbons 
and low optical albedos. As shown in Table~\ref{tab:predictions}, we find a substantial increase in hydrocarbon concentrations, compared to the solar case. For example, C$_2$H$_2$ and C$_2$H$_4$ are both enhanced by over six orders of magnitude each, and HCN is enhanced by over three orders of magnitude. Such enhanced concentrations of hydrocarbons might naturally lead to soot formation in CRP atmospheres. While the hydrocarbons C$_2$H$_2$, C$_2$H$_4$, and HCN might be observable with high resolution spectroscopy, e.g. with {\it JWST}, the effects of soot formation can be manifested in the form of low optical albedos of CRP atmospheres. 

Substantial depletion is predicted for most oxygen-based compounds. As discussed in section~\ref{sec: atmos-inv}, 
a depletion in TiO and VO is expected; we find TiO and VO depleted by factors of $\sim380$ and $\sim650$,  respectively. Such a depletion is consistent with the lack of a strong thermal inversion in WASP-12b reported by Madhusudhan et al. (2011). And, while CO$_2$ is depleted by a factor of $\sim 190$, consistent with the upper-limit of Madhusudhan et al. (2011), the abundance of SO$_2$ is decreased by over five orders of magnitude. As expected, the differences are not as pronounced for compounds without C or O, such as NH$_3$, PH$_3$, S$_2$, H$_2$S. 

 \begin{table}
\caption[]{Predicted Volume Mixing Ratios of Species in the atmosphere of WASP-12b for different C/O ratios}
\begin{center}
\begin{tabular}{lcc}
\hline
\hline
\noalign{\smallskip}
Species	&  Solar	& C/O = 1.1 \\	
\noalign{\smallskip}
\hline
\noalign{\smallskip}
H$_2$O	&	2.85E-04	&	4.13E-07	\\
CO	         &	4.02E-04	&	1.47E-03	\\
CH$_4$	&	6.16E-10	&	1.55E-06	\\
CO$_2$	&	2.12E-08	&	1.12E-10	\\
NH$_3$	&	4.01E-08	&	8.01E-08	\\
TiO	         &	5.90E-08	&	1.57E-10	\\
VO	         &	9.95E-10	&	1.53E-12	\\
C$_2$H$_2$ &	5.85E-12	&	3.72E-05	\\
C$_2$H$_4$ &	1.08E-15	&	6.85E-09	\\
HCN	         &	1.22E-08	&	6.17E-05	\\
PH$_3$	&	2.41E-10	&	1.19E-09	\\
SO$_2$	&	2.83E-12	&	8.75E-18	\\
S$_2$	&	1.93E-09	&	4.16E-09	\\
H$_2$S	&	4.33E-06	&	6.36E-06	\\
\hline
\end{tabular}
\end{center}
\tablecomments{The mixing ratios were computed using the best-fit temperature profile of WASP-12b reported 
in Madhusudhan et al. (2011) and assuming a representative quench pressure of 1 bar (e.g. Madhusudhan \& Seager 2011), where the temperature is 2845 K.}
\label{tab:predictions}
\end{table}

\section{Summary and Discussion}
\label{sec:discussion}

We have presented a general study of the atmospheric properties and interior abundances of carbon-rich giant (CRG) planets, defined as planets with C/O $\ge 1$. The first CRG atmosphere was recently discovered for the hot Jupiter WASP-12b, motivating our work. We explore a wide region of atmospheric phase space, ranging from the cooler end of known transiting giant planets ($T_{\rm eq} \sim 700$ K) to hot Jupiters and hot Neptunes at  $T_{\rm eq} \sim 2000$ K. We study the atmospheric concentrations of the dominant species (H$_2$O, CO, and CH$_4$) over a large grid in pressure-temperature space and the possibility of thermal inversions in CRG atmospheres, and we report observable spectroscopic features of such atmospheres. Based on the inferred C/H and O/H abundances for WASP-12b, the only CRG atmosphere known, we place constraints on the formation conditions of the planet and the elemental abundances in the envelope, using a core accretion model. 

CRGs probe a unique region in composition space. A C/O $\geq$ 1 causes distinct changes in the atmospheric molecular composition of C and O based species compared to solar abundances. At high temperatures (T $\gtrsim  1400$ K), and assuming solar abundances, CO and H$_2$O are the dominant C and O bearing species and CH$_4$ is a minor species. For an atmosphere with a C/O = 1, however, most of the oxygen is present in CO, H$_2$O is depleted by over 10 - 100 times compared to the solar case, and CH$_4$ is enhanced by over 10 - 100 times. Such a drastic difference in composition causes observable signatures in atmospheric spectra of CRGs that are detectable by existing instruments. For example, as evidenced in the case of WASP-12b (Madhusudhan et al. 2011), the low H$_2$O and high CH$_4$ are manifested as weak absorption in the numerous spectral bands of H$_2$O in the near to mid infrared, and strong absorption in the CH$_4$ bands at 3-4 $\micron$ and 7-8 $\micron$.  The differences between CRG spectra from solar abundance models are less evident at low temperatures ($T_{\rm eq} \le 800$K). At these temperatures, both carbon-rich and solar abundance models yield CH$_4$ and H$_2$O as the dominant carbon and oxygen bearing species, thereby making their atmospheres less distinguishable in chemical equilibrium. Non-equilibrium chemistry in the form of vertical mixing of CO alone is unlikely to 
cause significant difference between high and low C/O spectra. However, detailed photochemical calculations might reveal 
new pathways. For instance, if methane were destroyed and CO enhanced (Madhusudhan \& Seager 2011), the observable concentrations of H$_2$O and CO, along with other products, might be used to estimate the C/O ratio of the atmosphere. 

Our results show that even highly irradiated CRGs are not likely to host thermal inversions due to TiO and VO absorption in their atmospheres. Thermal inversions in hot Jupiters have traditionally been suggested to form due to gaseous TiO and VO which act as strong absorbers of incident visible light. We find that the C/O ratio strongly affects the abundance of TiO/VO available to form thermal inversions. A C/O = 1, for example, yields TiO and VO abundances that are $\sim$100 times lower than those obtained with the solar abundances that are typically assumed in the literature (Fortney et al. 2008). Such a depletion is adequate to rule out thermal inversions due to TiO and VO even in the most highly irradiated hot Jupiters, such as WASP-12b. However, the TiO and VO abundances may also be affected by other factors such as gravitational settling (Spiegel et al. 2009) and, potentially, be correlated with stellar activity (Knutson et al. 2010).

Based on the observed atmospheric C and O abundances, we place constraints on the formation conditions of WASP-12b in the protoplanetary disk. We assume a core accretion model with cold formation conditions ($T < 30$ K), which is the canonical model for Jupiter's formation in the solar nebula (Mousis et al. 2009a). We find that it is not possible to reproduce the C/O $\ge$ 1 observed in WASP-12b by the accretion of planetesimals formed in a disk whose initial gas phase elemental composition is similar to that of the parent star (C/O = 0.44). In order to reproduce the observed C/O ratio in the planet, one needs to invoke a primordial oxygen abundance in the disk which is depleted by a factor of $\sim 0.4$ compared to that of the parent star, with the exact value contingent on the volatile-to-silicate fraction. This formation scenario with a depleted O/H in the disk explains the entire range of C/H and O/H constraints obtained from observations of WASP-12b.

Current constraints on the atmospheric C/H and O/H ratios in WASP-12b are inadequate to decide between the alternative formation  scenarios. In our cold formation scenario, all the volatiles are in condensed form, implying that the heavy element abundances in the planetary envelope result primarily from planetesimal accretion. If, on the other hand, hotter conditions prevailed such that most of the oxygen were condensed in H$_2$O, and/or sequestered in rocks, whereas carbon-based volatiles such as CO and CH$_4$ were in gas phase in the disk, it might have been possible to obtain a C/O $\sim$ 1 in the planetary envelope due to gas accretion. Another possibility is that the planetesimals that trapped oxygen in their rocks or, at $T \sim $ 150 K, in the form of water ice, had adequate time to migrate inward into the disk and did not take part in the accretion onto the planet, leading to an oxygen-depleted gas accretion on to the forming planet. These scenarios, however, require that C/H ratios in the planetary envelope be close to the stellar C/H, and hence may not produce the super-stellar C/H ratios that are also consistent with the data for WASP-12b. These scenarios are inconsistent as well with the super-solar elemental abundances observed in Jupiter in the solar system. Stronger constraints on the C/H and O/H ratios in the atmosphere of WASP-12b in the future might help validate this scenario.

It is also possible that other processes, that occurred either during the formation or evolution of the planet, might have altered the C/O ratio in the atmosphere of WASP-12b. In particular, it is possible that a larger fraction of oxygen is sequestrated in species too refractory to be found in the atmosphere, or because the atmosphere itself is isolated from the interior by a radiative zone. This argument has already been invoked for the case of HD189733b (Mousis et al. 2009a; Mousis et al. 2011). WASP-12b is even more strongly irradiated than HD 189733b because of its closer proximity to the parent star. This can potentially lead to the development in the envelope of an outer radiative zone extending down to the kilobar pressure level. In this context, differential settling, resulting from the combination of gravity and irradiation effects, might take place inside the atmosphere, thus altering the volatile abundances in the upper layers (Baraffe et al. 2010) and leading to an atmospheric C/O ratio different from the interior. 

\acknowledgements{This study was supported in part by NASA grant NNX07AG80G. NM also acknowledges support through JPL/Spitzer Agreements 1328092, 1348668, and 1312647. O.M. acknowledges support from CNES. Most of the work was done while JIL was a Visiting Professor at the University of Rome ``Tor Vergata", and his contribution was financed within the scope of the Italian program ``Incentivazione alla mobilita' di studiosi straineri e italiani residenti all'estero." NM thanks Marc Kuchner, Sara Seager, Drake Deming, Adam Burrows, Julianne Moses, Ruth Murray-Clay, Dimitar Sasselov,  Aki Roberge, Adam Showman, Joe Harrington, Geoff Marcy, and Erik Petigura for helpful discussions.}

\end{document}